\def   \ni {\noindent}

\def   \ssk {\vskip  5truept}

\def   \bsk {\vskip 15truept}
 
\def   \newpage {\vfill\eject}
\def   \newline {\hfil\break}

\def\Ep{E_{\rm p}}

\def\E0{E_{\rm 0}}
\def\P0{\Phi_{\rm 0}}
\def\Ep0{E_{\rm p0}}

\documentstyle[epsfig]{article}
\begin{document}

\hsize 5truein
\vsize 8truein
\font\abstract=cmr8
\font\keywords=cmr8
\font\caption=cmr8
\font\references=cmr8
\font\text=cmr10
\font\affiliation=cmssi10
\font\author=cmss10
\font\mc=cmss8
\font\title=cmssbx10 scaled\magstep2
\font\alcit=cmti7 scaled\magstephalf
\font\alcin=cmr6 
\font\ita=cmti8
\font\mma=cmr8
\def\ref{\par\noindent\hangindent 15pt}
\null


\title{\ni SMOOTHLY BROKEN POWER LAW SPECTRA\\
OF GAMMA-RAY BURSTS}                     

\bsk \bsk
\author{\ni F.~Ryde}                       
\bsk
\affiliation{Stockholm Observatory, SE-133 36 Saltsj\"obaden, Sweden
}                                                
\bsk
\baselineskip = 12pt

\abstract{ABSTRACT \ni

A five-parameter expression for a smoothly broken power law is
presented. It is used to fit Gamma-Ray Burst spectra
observed by BATSE. 
The function is compared to previously
used four-parameter functions, such as a sharply broken power law and
the Band et~al. (1993) function.
The presented function exists as a WINGSPAN routine at
http://www.astro.su.se/$\sim$felix/wing.html
 }                                                    
\bsk
\baselineskip = 12pt

\bsk
\baselineskip = 12pt


\text{\ni 1. INTRODUCTION
\ssk
\ni

Studies of the spectral continuum of Gamma-Ray Bursts (GRBs) can give
 important clues for the
understanding of the underlying emission processes responsible
for the observed radiation. A number of studies have been done (Schaefer 
et al. 1992, 1994, Band et al. 1993 , Ford et al. 1994).
 The burst emission is of a non-thermal character and most of the
energy is released 
in the gamma-ray band (0.1 -- 1 MeV). 
Spectra from missions prior to the  
{\it Compton Gamma-Ray Observatory (CGRO)} with its Burst and 
Transient Source Experiment (BATSE) 
were often fitted with an ``optically thin thermal bremsstrahlung''
photon spectrum, $N _{\rm E} (E)\propto E^{-1} e^{-E/E_{0}}$, 
combined with a power law at larger energies. 
Schaefer et~al. also used a broken power law in fitting BATSE bursts.
Later, Band et~al. showed that nearly all time-integrated BATSE spectra were
well fitted by the four-parameter Band function consisting of 
two power laws, 
smoothly joined by an exponential roll-over. Below, we introduce a 
five parameter function, which consists of two
power laws smoothly joined together. The sharpness of the turn-over is
determined by the fifth parameter.

\bsk
\ni 2.1  SMOOTHLY BROKEN POWER LAW
\ssk
\ni

To model the time-integrated, background-subtracted photon count spectra,
we want a function describing  a 
smoothly broken power law, having a low energy power law with spectral 
index $\alpha$,  smoothly transferred into a high energy power law with
spectral index $\beta$. We follow the idea in Preece et al. (1996) and
define the following properties:
its logarithmic derivative varies  smoothly from $\alpha$ 
to $\beta$, with the transition described by a hyperbolic tangent
 function.
\newpage
\noindent
\begin{equation}
\frac{d \log N_{\rm E}}{d \log E} = \xi \tanh \left(\frac{\log(E/E_{\rm 0})}{\delta}\right) + \phi,
\end{equation} 
where $N_{\rm E}$ is the photon flux, $E$ is the photon energy,
$E_{\rm 0}$ is the break energy, 
$\delta$ is
the width of the transition, $\xi = (\beta - \alpha)/2$, and $\phi = (\beta +
\alpha)/2$. 
Integrating (1) gives 
\begin{equation}
N_{\rm E} (E) = C 
\left( \frac{E}{E_{\rm n}} \right) ^{\phi} 
\left[ \frac
{{\rm cosh}\left(\frac{{\rm log}(E/E_{\rm 0})}{\delta}\right)} 
{{\rm cosh}\left(\frac{{\rm log}(E_{\rm n}/E_{\rm 0})}{\delta}\right)}
\right]
^{\xi \delta {\rm ln}10},
\end{equation}
\noindent
where the width of the transition, in linear energy, is 
$\Delta E = E_{\rm 0}(10^\delta - 1)$.
The normalisation is set by the constant $C$, which is the value of 
$N _{\rm E} (E = E_n)$.
The function is described by four parameters besides the normalisation; 
the two power law indices, $\alpha$ and $\beta$, the break energy,
$E_{\rm 0}$, and the energy range over which the spectrum 
changes from one power law to the 
other, $\delta$ or $\Delta E$. The advantage of this function over the  
Band  function, for instance, which is described by four parameters, the 
normalisation, $\alpha$, 
$\beta$, and a parameter related to the break energy, is that the width 
of the transition region can be specified.

\begin{table*}
\caption[Table 1.]{ The bursts sample. The bursts are identified by their date
and trigger number (see the Current BATSE
Gamma-Ray Burst Catalogue, Meegan et~al.). $\Delta t$ and $\Delta$Energy
are  the chosen time and energy ranges used in the study.} 
\begin{flushleft}
\begin{tabular}{lccc |lccc}  
\\
\hline
\\
Date + &LAD &$\Delta t$ &$\Delta$Energy &Date + &LAD  &
$\Delta t$ &$\Delta$Energy
\\ 
Trigger No. &  & (s) &(keV) &Trigger No. &   & (s) &(keV)
\\
\\
\hline
\\
910430(\# 130)&6 &0-19 &29-1799 &940217(\# 2831) &0  &1-35 &33-1956 \\
910503(\# 143)&6 &0-10  &30-1766 &950403(\# 3491) &3 &0-23 &30-1649 \\
911127(\# 1122)&1 &0-38 &29-1774 &970420(\# 6198) &4 &1-13  &27-1950\\
920210(\# 1385)&5 &0-49&29-1988 &980125(\# 6581) &0 &45-59&24-1980\\
930506(\# 2329)&3 &0-14 &30-1756 &980203(\# 6587) &1 &0-34&25-1855\\
\\ 
\hline 
\end{tabular}
\end{flushleft}
\end{table*}
\begin{table*}
\caption[Table 2.]{ The results of the modelling of the bursts. $A$ is the
normalisation of the spectrum, $\alpha$ and $\beta$ are the low and
high energy power law photon indices, $E_{\rm 0}$ is the break
energy, and $\Delta E$
is the size of the turn-over spectral range. The goodness 
of fit is given by the $\chi
^2$-value together with the number of degrees of freedom, dof.
See the text for details.}
\begin{flushleft}
\begin{tabular}{llllllll}  
\\
\hline
\\
GRB &Model&$A / 10^{-2}$&$\alpha$&$E_{\rm 0}$&$\beta$&$\Delta E$&$\chi^2/$dof \\ 
\# & &(/cm$^{2}$/s/keV)& &(keV)&&(keV) &\\
\\
\hline
\\
130 &BPL  &0.97$\pm$0.01 &$-$1.46$\pm$0.02 &227$\pm$15 &$-$2.33$\pm$0.08 & &118/110 \\
  &Band &1.29$\pm$0.05 &$-$1.21$\pm$0.05 &395$\pm$50 &$-$2.5$\pm$0.2  &&128/110 \\
  &SBPL &0.97$\pm$0.01 &$-$1.45$\pm$0.02 &230$\pm$20 &$-$2.35$\pm$0.10
 &30$^{+60} _{-30}$ &118/109 \\

143  &BPL  &9.40$\pm$0.03 &$-$1.06$\pm$0.01 &382$\pm$7 &$-$2.09$\pm$0.02  & &236/109 \\
     &Band &10.9$\pm$0.07 &$-$0.91$\pm$0.01 &785$\pm$27 &$-$2.8$\pm$0.3  & &428/109 \\
     &SBPL &9.41$\pm$0.03 &$-$1.06$\pm$0.01&405$\pm$13 &$-$2.15$\pm$0.04&110$\pm$37 &229/108 \\

1122 &BPL  &1.38$\pm$0.01 &$-$1.48$\pm$0.02 &135$\pm$3 &$-$2.38$\pm$0.03 & &112/108 \\
     &Band &2.57$\pm$0.10 &$-$1.05$\pm$0.03 &163$\pm$9 &$-$2.51$\pm$0.06  & &84/108 \\
     &SBPL &1.40$\pm$0.01&$-$1.35$\pm$0.05 &143$\pm$9&$-$2.58$\pm$0.10  &112$\pm$40 &83/107 \\

1385 &BPL  &0.853$\pm$0.005&$-$1.02$\pm$0.01 &285$\pm$8&$-$2.23$\pm$0.04 & &180/113 \\
     &Band &1.14$\pm$0.02 &$-$0.73$\pm$0.02 &390$\pm$20
&$-$2.8$\pm$0.2  & & 127/113\\
     &SBPL &0.87$\pm$0.01 &$-$0.90$\pm$0.04 &425$\pm$80&$-$2.9$\pm$0.3  &610$\pm$310&122/112 \\

2329 &BPL  &8.30$\pm$0.02&$-$1.14$\pm$0.01 &300$\pm$9&$-$1.68$\pm$0.01& &161/108 \\
     &Band &9.38$\pm$0.07 &$-$1.03$\pm$0.01 &930$\pm$50 &$-$1.75$\pm$0.02  & &150/108 \\
     &SBPL &8.36$\pm$0.03 &$-$1.12$\pm$0.01&322$\pm$19&$-$1.73$\pm$0.03 &230$\pm$75&132/107 \\

2831 &BPL  &1.31$\pm$0.02 &$-$0.86$\pm$0.02 &424$\pm$13 &$-$2.07$\pm$0.03  & &166/112 \\
     &Band &1.55$\pm$0.02 &$-$0.61$\pm$0.03&635$\pm$35 &$-$3.0$\pm$0.4 & &122/112 \\
     &SBPL &1.32$\pm$0.02 &$-$0.75$\pm$0.06&760$\pm$210&$-$3.1$\pm$0.6 &1270$\pm$880 &118/111 \\

3491 &BPL  &4.08$\pm$0.02 &$-$1.65$\pm$0.01 &148$\pm$4 &$-$2.20$\pm$0.02  & &248/106 \\
     &Band &5.28$\pm$0.09 &$-$1.47$\pm$0.02&390$\pm$20 &$-$2.38$\pm$0.05 & &234/106 \\
     &SBPL &4.09$\pm$0.02 &$-$1.62$\pm$0.02 &183$\pm$13&$-$2.37$\pm$0.06 &145$\pm$50 &225/105 \\

6198 &BPL  &8.55$\pm$0.03 &$-$1.29$\pm$0.01 &198$\pm$3 &$-$2.23$\pm$0.02 & &277/113 \\
     &Band &11.90$\pm$0.15 &$-$1.04$\pm$0.01&325$\pm$10 &$-$2.53$\pm$0.05& &201/113 \\
     &SBPL &8.68$\pm$0.04 &$-$1.24$\pm$0.02&246$\pm$12 &$-$2.51$\pm$0.06  &204$\pm$43 &175/112 \\

6581 &BPL  &2.86$\pm$0.01 &$-$1.56$\pm$0.01 &195$\pm$11 &$-$2.07$\pm$0.04  & &112/114 \\
     &Band &3.39$\pm$0.07&$-$1.43$\pm$0.02 &625$\pm$70 &$-$2.20$\pm$0.10 & &121/114 \\
     &SBPL &2.87$\pm$0.02 &$-$1.55$\pm$0.01 &212$\pm$20&$-$2.12$\pm$0.07 &65$\pm$55 &111/113 \\

6587 &BPL  &5.35$\pm$0.01 &$-$1.26$\pm$0.01 &194$\pm$3 &$-$2.08$\pm$0.01  & &356/111 \\
     &Band &7.35$\pm$0.08 &$-$1.01$\pm$0.01 &338$\pm$10 &$-$2.25$\pm$0.03 & &261/111 \\
     &SBPL &5.44$\pm$0.02 &$-$1.20$\pm$0.02 &220$\pm$7 &$-$2.24$\pm$0.04  &170$\pm$30 &241/110 \\
\hline 
\end{tabular}
\end{flushleft}
\end{table*}

\bsk
\ni 3. GRB SPECTRA OBSERVED BY BATSE
\ssk
\ni

We have studied a sample of 10 strong GRBs, for which the Smoothly
Broken Power Law (SBPL) above  
gives reasonable fits.
They are used to illustrate the comparison between 
the Band model and the SBPL model.
In Table 1, the set of bursts is displayed, together with the
time and energy range used and the number of 
the Large Area Detector (LAD) used.
The following three models were used:\\
(1) A sharply Broken Power Law (BPL) consisting of two power
laws connected at the break energy with three parameters besides the 
normalisation:  the photon indices of the low and high energy power
laws, $\alpha$, $\beta$, and the break energy, $E_{0}$.\\ 
(2) The Band et~al. (1993) function (Band) with three parameters
besides the normalisation: $\alpha$,
$E_{\rm 0}$, $\beta$. The spectrum below $E = (\alpha - \beta) E_{\rm 0}$ 
 is an exponentially cut off power law, $N_{\rm E}(E) \propto
E^{\alpha}{\rm e}^{-E/E_{\rm 0}}$. A second power law, 
$N_{\rm E}(E) \propto E^{\beta}$,
is connected to this  function above this energy.
Note that $E_{\rm peak} \equiv (2 + \alpha)E_{\rm 0}$ is usually used to
characterise the break in the Band function.
\\ 
(3) The Smoothly Broken Power Law (SBPL) given by Equation (2)
with its four parameters besides the normalisation.\\
The results of the fitting are shown in Table 2. 
The general trends of the sample studied are the following. \\
(1) The Band model gives the
largest break energy (except for \# 2831, which, however, has large errors).\\
(2) The SBPL model has a somewhat larger break energy than the BPL model.\\
(3) The low energy power law in the Band model is harder than the one
in the  SBPL model, which is marginally
harder than the one in the BPL model.\\
(4) The high energy power law in  the BPL model is the hardest and in 
the  Band  model it is the softest.
 
The main difference between the SBPL model and the Band model is the extra
variable parameter describing the energy range of the changing of the power law
index. As bursts can have an actual
turn-over that is sharper
than the one given by the fixed exponential turn-over in the Band
function, which is determined solely by $E_{\rm 0}$, the resulting
fits in Table 2 differ.
The smaller the ratio between $\Delta E$ and $E_{\rm 0}$ in the SBPL model
is, the larger the difference is between $E_{\rm 0}$ in the Band
model and in the SBPL model. The best examples of this 
trend are \# 6581 and \# 143 compared to \# 1385 and \# 1122.
The smooth turn-over in the Band model allows, among other things, the low 
energy power law to be harder than what is found by the BPL model.
However, extra parameters are not always advantageous. In some cases, not shown
in Table 2, the $\Delta E$ parameter in the SBPL model cannot be constrained.

\bsk
\baselineskip = 12pt
{\abstract \ni ACKNOWLEDGEMENTS

Thanks are due to Roland Svensson and Jerry Bonnell (GROSSC). Financial
support was received from the Swedish National Space Board and a
Nordic Project grant at Nordita.

}

\bsk
\baselineskip = 12pt


{\references \ni REFERENCES
\ssk

\ref
Band, D.L., et~al. 1993, ApJ, 413, 281

\ref
Ford, L.A., et~al. 1995, ApJ, 439, 307

\ref
Meegan, C.A., et~al. 1998, http://www.batse.msfc.nasa.gov/data/grb/catalog/

\ref
Preece, R.D., Briggs, M.S., Mallozzi, R.S., \& Brock, M.N. 1996, 
WINGSPAN v 4.4 manual

\ref
Schaefer, B. et~al.  1992, ApJ,   393, L51

\ref
Schaefer, B. et~al.  1994, ApJS, 92, 285   

}                      

\end{document}